%% file: main.tex
\documentclass[twocolumn,10pt,aps,nofootinbib,showkeys,superscriptaddress,prc]{revtex4-2}
\usepackage{orcidlink}
\usepackage{physics}
\usepackage{qcircuit}

\usepackage[utf8]{inputenc}
\usepackage[T1]{fontenc}
\usepackage{newunicodechar}
\usepackage{wasysym} 
\newunicodechar{◑}{\LEFTcircle}
\newcommand{\halfctrl}{\raisebox{0.1em}{\scalebox{0.8}{\LEFTcircle}}}

\newcommand{\ecut}{\varepsilon_\text{cut}}
\newcommand{\cost}[1]{\text{Cost}\!\left(\text{#1}\right)}

\begin{document}
\bibliographystyle{ieeetr}

\title{Systematic improvement of the quantum approximate optimisation ansatz for combinatorial optimisation using quantum subspace expansion}

\author{Yann Beaujeault-Taudi\`ere \orcidlink{0000-0002-7317-1287}}
\email{yann.beaujeault-t@protonmail.com}
\noaffiliation

\date{June 19, 2025}

\begin{abstract}
The quantum approximate optimisation ansatz (QAOA) is one of the flagship algorithms used to tackle combinatorial optimisation on graphs problems using a quantum computer, and is considered a strong candidate for early fault-tolerant advantage. In this work, I study the enhancement of the QAOA with a generator coordinate method (GCM), and achieve systematic performances improvements in the approximation ratio and fidelity for the maximal independent set on Erd\"os-Rényi graphs. The cost-to-solution of the present method and the QAOA are compared by analysing the number of logical CNOT and $T$ gates required for either algorithm. Extrapolating on the numerical results obtained, it is estimated that for this specific problem and setup, the approach surpasses QAOA for graphs of size greater than 75 using as little as eight trial states. The potential of the method for other combinatorial optimisation problems is briefly discussed.
\end{abstract}

\maketitle

\section{Introduction}

Among the myriad fundamental and applied use cases that quantum computing is expected to disrupt, combinatorial optimisation on graphs stands a peculiar role, finding applications across several fields, such as computational biochemistry~\cite{bhattacharya2005graph,mendes2005protein} and genetics~\cite{aluru2006handbook}, statistical mechanics~\cite{stauffer1994percolation,weigt2001hardspherefluidsolid}, resource allocation~\cite{korte2006combinatorial,hillier2005introduction} and logistics~\cite{button2010handbook,toth2002vehicle}, or circuit design~\cite{nebel2008vlsi,sherwani1999algorithms}. The possibility to achieve quantum advantage for solving NP-hard instances has been the focus of extensive experimental and theoretical research in the recent years, within the scope of both analogue ~\cite{johnson_quantum_2011,ebadi_quantum_2022,dalyac_exploring_2023,kim_quantum_2024,leclerc_implementing_2024} and digital ~\cite{Amaro2021,Harrigan2021,Jain2022,Zhou2022,Ponce2023,Dupont2023} approaches. The quantum approximate optimisation ansatz (QAOA)~\cite{farhi_quantum_2014,Blekos_2024} has emerged as one of the most promising variational algorithms to tackle graph problems. Several variations have been put forth, aiming at reducing the circuit depth by using more sophisticated mixers~\cite{zhu_adaptive_2020,herrman_multi-angle_2022}, or the number of evaluations of the cost function during the optimisation~\cite{lee_iterative_2023}. The existing flavours of QAOA all add  refinements of varying complexity to the original ansatz.

In this work, an alternative and complementary route is followed, combining the QAOA ansatz with a generator coordinate method (GCM), which is a standard tool for approximating the solution to eigenvalues problems on classical processors, and which has more recently been adapted to quantum computers for the study of strongly correlated electronic and nuclear systems~\cite{zheng_quantum_2022,jamet_quantum_2022,beaujeault-taudiere_solving_2023}. Due to the strong link between combinatorial optimisation and many-body Hamiltonians, it is natural to expect that tools can be ported from one domain of applications to the other. Here, I restrict the GCM to its quantum subspace expansion (QSE) implementation, that is, a configuration-mixing scheme between real-time-evolved states built on top of a single wavefunction.

As an illustration, I focus on the maximal independent set (MIS) problem~\cite{douglas2000introduction,gross2018graph}, which, due to its NP-hardness, real-world use cases, and mapping to the native Hamiltonian of neutral atom platforms~\cite{Henriet_2020}, has been the subject of a vast body of studies~\cite{Pichler2018,Kim2021,ebadi_quantum_2022,Nguyen2023_quantum,kim_quantum_2024}. Additionnaly, I introduce a simple metric to identify whether performing the quantum subspace expansion step on top of an initial state provided by the QAOA improves the cost-to-solution ratio. I apply the QAOA-plus-QSE to Erd\"os-Rényi graphs of average density $\rho=1/2$, and extrapolate the critical graph size for which the method outperforms QAOA for this specific problem.

This paper is organised as follows. Section \ref{sec:mis}, \ref{sec:qaoa}, \ref{sec:gcm} respectively present the general aspects of the MIS problem, of the QAOA, and of the GCM. The evaluation of the so-called Hamiltonian kernels, which represent the most challenging quantity to evaluate on a quantum processing unit (QPU), is discussed in section \ref{sec:kernels_computation}; three different procedures are analysed. In particular, a resource evaluation in terms of CNOT and $T$ gates is given. Section \ref{sec:when_is_gcm} defines a simplified metric that can be used to determine whether applying the QSE protocol on top of a circuit optimised by QAOA leads to an improvement in the time-to-solution. Section \ref{sec:gcm_results} presents numerical results.

\section{Maximal independent set}\label{sec:mis}

The maximum independent set is a classic problem of graph theory~\cite{douglas2000introduction,gross2018graph}. Given a graph $G=(V,E)$, where $V$ is the set of vertices and $E$ is the set of edges, the objective is to find the largest set of independent vertices, that is, the largest set where no two vertices are linked by an edge. The MIS problem is generally NP-hard, which means that there is no known classical algorithm that can find the optimal solution in polynomial time. There exist however efficient heuristic algorithms that can get close to the optimal solution, provided the MIS in question belongs to the appropriate complexity class~\cite{dalyac_exploring_2023}. The MIS problem can be formalised as finding the minimum of the cost Hamiltonian
\begin{align}\label{eq:cost_hamiltonian}
    \hat{H}_C = -\sum_{v_i\in V} \hat{n}_i + \sum_{(i,j) \in E} \hat{n}_i \hat{n}_j,
\end{align}
where the number operator $\hat{n}_i=(1-Z_i)/2$ corresponds to rejecting ($\langle \hat{n}_i \rangle = 0$) or including ($\langle \hat{n}_i \rangle = 1$) the node $i$ in the solution. Expanding \eqref{eq:cost_hamiltonian} in terms of the Pauli matrices $Z_i$ puts the Hamiltonian in a form suited for quantum simulation. Solving the MIS using QPUs has mostly been studied from the angle of adiabatic evolution~\cite{kim_quantum_2024,zhao2024quantumhamiltonianalgorithmsmaximum}, quantum annealing~\cite{Kim2021,leclerc_implementing_2024}, and QAOA~\cite{brady2023iterativequantumalgorithmsmaximum,wybo2024missingpuzzlepiecesperformance,ni2025progressivequantumalgorithmmaximum,xu2025qaoaparametertransferabilitymaximum}. Since the GCM is a digital algorithm, I narrow down the present study to the QAOA, which I succinctly review in the next section.

\section{Quantum approximate optimisation ansatz}\label{sec:qaoa}

The quantum approximate optimisation ansatz (QAOA)~\cite{farhi_quantum_2014,Blekos_2024} is one of the leading algorithms for solving combinatorial optimisation problems~\cite{Blekos_2024}. In its original version, QAOA makes use of two non-commuting operators $\hat{H}_C$ and $\hat{H}_M$, called the cost and mixer Hamiltonians, respectively. Together, they define the alternating parametric unitary of depth $L$
\begin{align}\label{eq:def_U}
    U(\boldsymbol{\gamma},\boldsymbol{\beta}) = \prod^L_{\ell=1} e^{-i\beta_\ell \hat{H}_M} e^{-i\gamma_\ell\hat{H}_C},
\end{align}
and one seeks to minimise the following expectation value:
\begin{align}\label{eq:def_QAOA}
    C(\boldsymbol{\gamma},\boldsymbol{\beta}) 
    &= \mel{\phi}{U^\dag(\boldsymbol{\gamma},\boldsymbol{\beta}) \hat{H}_C U(\boldsymbol{\gamma},\boldsymbol{\beta})}{\phi} \nonumber\\
    &\equiv \mel{\Phi_0}{\hat{H}_C}{\Phi_0},
\end{align}
In order to lower the energy, the mixer Hamiltonian should not commute with $\hat{H}_C$. It should furthermore allow reaching the ground state of the cost Hamiltonian, which can be formalised as the requirement that the nested commutators of the cost and mixer produce an algebra whose exponential map contains at least one path transforming $\ket{\phi}$ into the proper ground state. These two conditions can be fulfilled regardless of the initial state by the simple yet efficient choice of a coherent rotation mixer:
\begin{align}\label{eq:coherent_mixer}
    \hat{H}_M = -\sum^{N-1}_{i=0} X_i,
\end{align}
where $X_i$ is the Pauli $X$ operator acting on qubit $i$. 

As the state that minimises $C$ is a computational state\footnote{In the case where more than one computational state minimises $C$, any linear combinations of them has the same energy and is thus also a valid solution.}, $\ket{\phi}$ is often chosen to be the uniform superposition $\ket{+}^{\otimes N}$, which guarantees a non-zero overlap with the unknown optimal of $C(\boldsymbol{\gamma},\boldsymbol{\beta})$. It is furthermore the ground state of the mixer \eqref{eq:coherent_mixer}, which allows connecting the QAOA to adiabatic evolution.

One of the main challenges of QAOA is the possible existence of barren plateaus (BP) in the cost function landscape, making the search for the ground state exponentially difficult with the number of nodes~\cite{McClean_2018_barren,Larocca_2022_diagnosing,Larocca_2025_barren}. The existence or absence of BP is intimately linked to the dimension of the dynamical Lie algebra (DLA) associated to the alternating layer ansatz~\cite{anand2022exploring,allcock2024dynamical,kazi2024analyzing}. In addition to the DLA arguments, which assumes the number of layers $L$ is infinite, BPs can emerge in finite-depth ansatzes if $L$ is sufficiently large, typically when $L$ is polynomial with respect to the number of nodes $N$. As finding the ground state of Ising Hamiltonians is generally NP-hard, it is unlikely that QAOA can reach the ground state in $\text{poly}(N)$ layers. This simple argument indicates a potential risk of encountering BPs when solving combinatorial optimisation tasks using QAOA. In order to mitigate or alleviate the vanishing of the cost function gradients, many strategies have been proposed, including warm-starting~\cite{tate_warm-started_2024}, optimising a small number of layers at a time~\cite{lee_iterative_2023}, or iteratively selecting the mixers among a pool~\cite{zhu_adaptive_2020,herrman_multi-angle_2022}. Another appealing strategy is to use QAOA as a starting point for quantum generator coordinate methods, as I present in the next section.

\section{Generator coordinate method}\label{sec:gcm}

The generator coordinate method (GCM) ~\cite{hill_nuclear_1953,griffin_collective_1957,Wa_Wong1975-oj,ring_nuclear_1980,bender_self-consistent_2003,verriere_time-dependent_2020} is a general method to obtain ground and excited states properties of a Hamiltonian. It is one of the standard tools of many-body quantum physics, with a long history of being used on classical computers to access static and dynamic properties of many-body Hamiltonians across several energy scales. This includes non-perturbative and perturbative phenomena alike, such as dynamical fission paths in atomic nuclei~\cite{Bertsch_2017,Bertsch_2022}, large-amplitude deformations modes~\cite{severyukhin_beyond_2006,Nikcic_2011,bally_symmetry_2012,Egido_2016,Zhao_2019}, and small amplitude vibrations~\cite{jancovici_collective_1964,Jiao_2019,marevic_quantum_2023}. The GCM approach has been recently applied on quantum computers in order to access properties of strongly coupled fermionic systems, such as the one-body Green's functions~\cite{jamet_quantum_2022,umeano2024quantumsubspaceexpansionapproach}, ground and excited states building on a unitary coupled clusters ansatz~\cite{zheng_quantum_2022,zheng_unleashed_2024} or coherent one-body rotations in the case of Hamiltonians possessing permutation invariance~\cite{beaujeault-taudiere_solving_2023}. The GCM is based upon expanding the eigenstates of the cost Hamiltonian on a set of (usually non-orthogonal) generating states 
$\{\ket{\chi_{k}(\boldsymbol{q}_{k})}\}_{k=0,\ldots,K-1}$ according to 
\begin{align}\label{eq:gcm_expansion}
    \ket{\Psi_j} = \sum^{K-1}_{k=0} f_{k,j} \ket{\chi_{k}(\boldsymbol{q}_{k})},
\end{align}
the parameters $\boldsymbol{q}_{k}$ being referred to as the generator coordinates. One has full freedom in choosing the generating states; for instance, this formulation encompasses real-time quantum subspace expansion (QSE)~\cite{parrish_quantum_2019,stair2019multireferencequantumkrylovalgorithm,lee2021variationalquantumsimulationchemical,jamet_quantum_2022} as the special case $\boldsymbol{q}_{k}=t_{k}$ and $\ket{\chi_{k}(t_k)}=e^{-i\hat{H}_Ct_{k}}\ket{\Phi_0}$\footnote{From a formal point of view, it is important that the dimension of the Hermitian operator used to define the generating manifold be as large as possible in order to eventually eliminate all excited states in the limit of large $K$. Otherwise, the Krylov subspace spanned by the generating states might be too low-dimensional to yield satisfactory approximations to the true eigenstates. In that respect, choosing the cost Hamiltonian itself as the generator of the trial states is optimal.}. Another possibility that is commonly employed for the description of atomic nuclei is to depart from the operator-based expansion, and generate the trial states using constrained optimisation, where the coordinates correspond to expectation values of deformation operators, whose selection is essentially based on phenomenological considerations~\cite{ring_nuclear_1980,bender_self-consistent_2003,marev2018} \footnote{When the initial state $\ket{\Phi_0}$ is a product state, this second approach is a special case of unitary evolution, corresponding to a Thouless transformation.}. In the most general case, the values of the coordinates can be optimised variationally, possibly for each target eigenstate. This is however rarely done in practice, for two different reasons. When the GCM is employed to restore symmetries, both the weights $\{f_{k,j}\}$ and the coordinates can be derived analytically~\cite{hamermesh_group_1962,lipkin_lie_1966,ring_nuclear_1980,bender_self-consistent_2003,bally_symmetry_2012}. Alternatively, when the GCM is used to grasp correlations that are not captured by the single-reference state $\ket{\Phi_0}$, the coordinates are fixed because of the prohibitive cost of the associated variational optimisation using classical computers. As an illustration, fully variational implementations of the GCM, that is, where not only the weights but also the states, are optimised, have been achieved only very recently in the context of nuclear physics~\cite{matsumoto_extension_2023}.

Given the expansion \eqref{eq:gcm_expansion}, approximate eigenstates of $\hat{H}_C$ can be obtained by applying the Rayleigh-Ritz variational principle~\cite{Ritz1909} to the cost function
\begin{align}\label{eq:cost_gcm}
    C^{(j)}_\text{GCM}(\{f_{k,0}\}) = \sum_{kk'} f^*_{k,j} f_{k',j} \mel{\chi_k}{\hat{H}_C}{\chi_{k'}},
\end{align}
with the normalisation
\begin{align}\label{eq:norm_gcm}
    \sum_{kk'} f^*_{k,j} f_{k',j} \braket{\chi_k}{\chi_{k'}} = 1,
\end{align}
leading to a generalised eigenvalue equation
\begin{align}\label{eq:gcm_hillwheeler}
    \mathcal{H}\ket{\Psi} = E\mathcal{S}\ket{\Psi},
\end{align}
where $\mathcal{H}$ and $\mathcal{S}$ are the so-called Hamiltonian and overlap kernels, whose elements read
\begin{align}
    \mathcal{H}_{kk'} &= \mel{\chi_k}{\hat{H}_C}{\chi_{k'}}, \label{eq:kernels_definition_H}\\
    \mathcal{S}_{kk'} &= \braket{\chi_k}{\chi_{k'}}. \label{eq:kernels_definition_S}
\end{align}
Due to the non-orthogonality of the generating states, the overlap matrix may contain small eigenvalues which spoil the numerical resolution of Eq.\eqref{eq:gcm_hillwheeler}, even when the matrix elements are computed exactly~\cite{bonche_analysis_1990,marev2018,epperly_theory_2023,beaujeault-taudiere_solving_2023}. This instability is exacerbated in the presence of noise. A simple prescription is to project out the states that are quasi-degenerate with the rest of the basis, by removing the eigenvalues that are smaller than a given threshold $\ecut$ (see App.~\ref{app:gcm_reencoding}. This has been used successfully in the nuclear physics context since over three decades~\cite{bonche_analysis_1990,marev2018}, delivering robust and reliable results for virtually any reasonable choice of the threshold. The convergence and stability properties of this approach were studied in detail in ~\cite{epperly_theory_2023}. 

Alternatively to solving \eqref{eq:gcm_hillwheeler} in the truncated basis, another procedure is to use the quantum state deflation method (QSD) proposed in ~\cite{higgott_variational_2019}, which allows determining the eigenstates iteratively by optimising the sequence of cost functions
\begin{align}\label{eq:qsd}
    C^{(j)}_\text{GCM} = &\mel{\Psi_j}{\hat{H}_C}{\Psi_j} - \mu_0 (1 - \braket{\Psi_j}{\Psi_j})^2 \nonumber\\
    &- \sum^{j-1}_{j'=0} \lambda_{j'} \abs{\braket{\Psi_{j'}}{\Psi_j}}^2.
\end{align}
Here, the Lagrange multipliers $\mu_0$ and $\{\lambda_{j'}\}$ respectively ensure the normalisation of the wave function and its orthogogonality with respect to the previously found eigenstates.  The diagonalisation-based and QSD-based methods differ only in the way the kernels are post-processed classically. It is noteworthy that the two methods are not strictly equivalent, as a consequence of the arbitrariness of the truncation energy of the diagonalisation method, and of the numerical values of Lagrange multipliers within the QSD. Nonetheless, they were found to yield similar spectra as long as the many-body kernels are determined to sufficient accuracy~\cite{beaujeault-taudiere_solving_2023}.

Another notable feature of the QSE is that performance guarantees can be derived. For instance, it can be shown that the weights can be chosen such that QSE on a reguler time grid implements a Gaussian energy filter or an imaginary time evolution, see App.~\ref{app:kernels_free_gcm}.

Finally, I mention that since the QAOA requires the breaking of at least one symmetry of $\hat{H}_C$ in order for the value of the cost function to decrease, the symmetry restoration capabilities of the GCM are particularly appealing in that case. This feature was illustrated in ~\cite{beaujeault-taudiere_solving_2023}, where a well-defined parity was automatically recovered in the eigenstates of a many-body Hamiltonian. Likewise, in the context of combinatorial optimisation on graphs, where the ground state is a computational state with well-defined (but unknown a priori) Hamming weight (also called particle number in many-body physics), the GCM is found to systematically increase the magnitude of the states with the correct Hamming weight.

\section{When is GCM a good FTQC algorithm?}\label{sec:when_is_gcm}

Whether an initial state algorithm should be employed rather than another is measured by whether it reduces the overall computational resources, that is, runtime, number of gates, and/or number of logical qubits required. Among these metrics, the number of gates plays the most critical role, as it essentially dictates the balance between space and time tradeoffs during the execution on the QPU. Focusing on this quantity alone is thus a good first first-order estimate of an algorithm's performances, and in the following discussion, I use the term ``cost'' to refer to the number of gates required to run the algorithms. 

The fidelity between the ground state(s) and the state prepared by a given algorithm $``X"$ is written $F^{(X)}$. The cost-to-fidelity ratio of the GCM is favourable if the following condition is met:
\begin{align}\label{eq:cost_to_fidelity}
    \frac{F^\text{(GCM)}}{F^\text{(QAOA)}} \gtrsim 4(K-1) \left(1+\frac{1}{L'}\right) \times f,
\end{align}
where $f$ is a scaling factor depending on which of the subcircuits introduced in Sec.~\ref{sec:kernels_computation} is used to extract the matrix elements of the kernels (see App.~\ref{app:gate_counts}), and $L'\leq L$ is the number of layers retained after the optimisation of the QAOA angles.

From \eqref{eq:cost_to_fidelity}, it can readily be argued that under some assumptions, there will always exist some critical graph size beyond which the GCM is always favourable over the plain QAOA, in the case of a fixed depth. Indeed, while there are no rigorous results on the practical scaling of the QAOA fidelity with the number of nodes, a simple empirical fit of the form $F^{(X)}(N)=\beta_X(1+e^{N \alpha_X})^{-1}$ captures the most salient expected features of the fidelity, namely its smooth decay from one to (almost) zero as $N$ grows, and its plateau effect for small $N$. As the left-hand side of \eqref{eq:cost_to_fidelity} grows exponentially, while the right-hand side scales polynomiallly, the GCM will deliver a better cost-to-fidelity ratio than QAOA for larger graphs. In Sec.~\ref{sec:gcm_results}, this simple metric is used to extrapolate the QAOA and GCM results in the special case of Erd\"os-Rényi graphs with fixed average density.

\section{Computation of the kernels}\label{sec:kernels_computation}

\begin{figure}[ht!]
    \begin{center}
        \resizebox{0.9\linewidth}{!}{\input{circuits_sketches/hadamard_test}}
    \end{center}
    \caption{Circuit for the evaluation of $A_{kk'}$ with Hadamard tests. The real part of $A_{kk'}$ is obtained by setting $\phi$ to zero. The imaginary part is obtained using $\phi=-\pi/2$, which corresponds to the modified Hadamard test. Both parts are recovered as the difference between the probabilities of measuring the ancillary qubit in the state zero and one.}
    \label{fig:hadamard_test}
\end{figure}
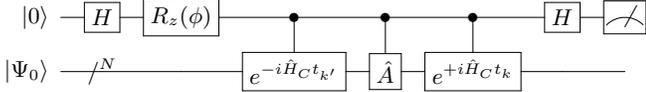

The matrix elements $A_{kk'}\equiv\mel{\Psi_{k}}{\hat{A}}{\Psi_{k'}}$ can be evaluated using two Hadamard tests, as shown in Fig.~\ref{fig:hadamard_test}. 
While evaluating $S_{kk'}$ poses little challenge, estimating the kernels $\mathcal{H}_{kk'}$ requires more care, as the cost Hamiltonian is not a unitary operator. This can be addressed in several ways:

\begin{itemize}
    \item By estimating the expectation values of the Pauli strings $\{Z_n,Z_nZ_{n'}\}$ directly. With this approach, $N/2$ two-qubits expectation values can be estimated simultaneously by grouping the operators into commuting sets. This approach furthermore avoids the use of extra CNOT and $T$ gates beyond the ones required to encode the states. Because $\mathcal{O}(\sqrt{\rho}N)$ rounds of measurements are performed, each with approximately $\sqrt{\rho}N$ terms evaluated, the total number of shots scales as $(\sqrt{\rho} N)^3$.
    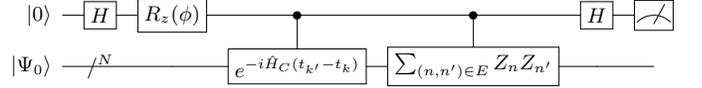
\begin{figure}[ht!]
        \begin{center}
            \resizebox{0.95\linewidth}{!}{\input{circuits_sketches/hadamard_Pauli}}
        \end{center}
        \caption{Circuit implementing the Hadamard test for estimating $\sum_{n,n'}\mel{\Psi_{k}}{Z_nZ_{n'}}{\Psi_{k'}}$. Ideally, all the sums run over distinct maximal bipartite matchings of the edges.
        }
        \label{fig:Hadamard_Pauli}
    \end{figure}
        
    \item By using real time evolution and forming linear combinations. In that case, one estimates $\mel{\Psi_{k}}{e^{-i\hat{H}_Ct_j}}{\Psi_{k'}}=\mathcal{S}_{kk'}-it_j\mathcal{H}_{kk'}+\mathcal{O}(t_j^2)$ for several propagation times $\{t_j\}_{j=1,\cdots,p}$. The kernels $\mathcal{S}_{kk'}, \mathcal{H}_{kk'}$ can then be recovered to precision $\mathcal{O}(t^p)$ by forming linear combinations of the $p$ expectation values. The principal disadvantage of this approach is that it requires large values of $p$ in order to estimate matrix elements to high precision, which in turns increases the final variance. The circuit used to estimate the expectation values is shown in Fig.~\ref{fig:Hadamard_RTE}.
    \begin{figure}[ht!]
        \begin{center}
            \resizebox{0.85\linewidth}{!}{\input{circuits_sketches/hadamard_RTE}}
        \end{center}
        \caption{Circuit implementing the Hadamard test plus RTE method for estimating $\langle e^{-i\hat{H}_C t_j} \rangle_{kk'} \equiv \mel{\Psi_{k}}{e^{-i\hat{H}_C t_j}}{\Psi_{k'}}$. The circuit is run for all values of $t_j$, yielding the expectation values $\{\langle e^{-i\hat{H}_C t_j} \rangle_{kk'}\}_{j=1,\cdots,p}$. These are eventually combined in a post-processing step, giving $\mathcal{H}_{kk'} = \sum^p_{j=1} a_j \langle e^{-i\hat{H}_C t_j} \rangle_{kk'} + \mathcal{O}(t^{p})$. The same matrix elements can be ued to extract $\mathbf{S}_{kk'}$.
        }
        \label{fig:Hadamard_RTE}
    \end{figure}
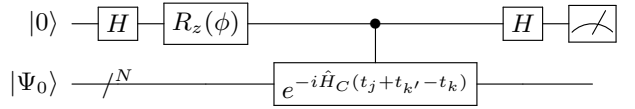
    
    \item By encoding $\hat{H}_C$ into a larger space using the linear combination of unitaries (LCU) approach~\cite{childs_hamiltonian_2012}. This approach involves $\mathcal{O}(\log_2(N))$ ancillary qubits. The LCU is a probabilistic algorithm, and close to the optimal solution, $p_\text{success} \sim (\log_2(N)/N)^4$.
    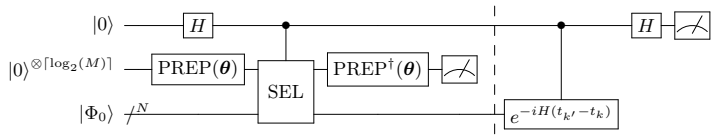
\begin{figure}[ht!]
    \begin{center}
        \resizebox{0.95\linewidth}{!}{\input{circuits_sketches/Hadamard_LCU}}
    \end{center}
    
    \caption{Circuit implementing the Hadamard test plus LCU method for estimating the kernels. The part of the circuit after the barrier is executed only if the ancilla register is measured in the ${0}^{\otimes \lceil \log_2(M)\rceil}$ state. Here, $M=N(N+1)/2$ is the total number of Pauli strings in the Hamiltonian. Hence, at leading order, $\lceil\log_2(M)\rceil\approx\lceil 2\log_2(N) \rceil$.}
    \label{fig:Hadamard_LCU}
\end{figure}
\end{itemize}

Which approach is optimal depends on several parameters, including the availability of ancillary qubits, the required precision to which the unitaries must be realised, which algorithm is used to decompose them on the elementary gate set, and the scaling of the off-diagonal moments $\mel{\Psi_{k}}{\hat{H}^p_C}{\Psi_{k'}}$. The three approaches are analysed and compared in more depth in App.~\ref{app:gate_counts}.

\section{Re-encoding the GCM states}\label{sec:states_reencoding}

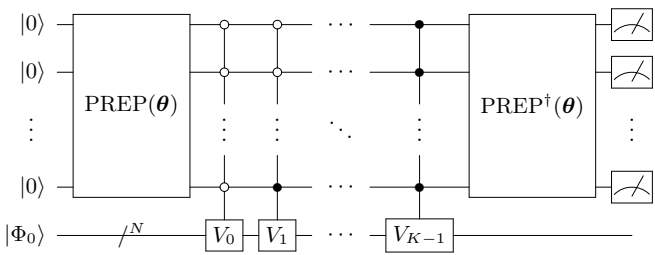
\begin{figure}[ht!]
    \begin{center}
        \resizebox{0.95\linewidth}{!}{\input{circuits_sketches/LCUv2}}
    \end{center}
    \caption{LCU circuit encoding the state $\ket{\Psi_j}=\sum^{K-1}_{k=0}f_{k,j}(\boldsymbol{\theta})V_k\ket{\Phi_0}$ into the state register. The encoding is successful if the $\lceil\log_2(K)\rceil$ ancillas are measured in the state $0$, and fails otherwise.}
    \label{fig:LCU}
\end{figure}

Once the weights $\{f_{k,j}\}$ are obtained, it is possible to encode the linear combination \eqref{eq:gcm_expansion} directly into the register using the LCU, as shown in Fig.~\ref{fig:LCU} and detailed in App.~\ref{app:gcm_reencoding}. The \texttt{PREP}$(\boldsymbol{\theta})$ unitary implements a  transformation whose entries are related to the weights according to (up to a permutation) $[\texttt{PREP}(\boldsymbol{\theta})]_{k0}=\sqrt{f_{k,0}(\boldsymbol{\theta})}$. This re-encoding step is required to recover the bitstring that minimises the cost function, and also offers the possibility to compute arbitrary expectation values $\mel{\Psi_j}{\hat{O}}{\Psi_j}$ without estimating all the corresponding kernels. The re-encoding of the state $\ket{\Psi_j}$ is successfull with probability
\begin{align}\label{eq:re-encoding_proba}
    p_0 = \frac{\sum_{kk'} f^*_{k,j}f_{k',j} \mathcal{S}_{kk'}}{\left(\sum_{k} \abs{f_{k,j}}\right)^2}.
\end{align}

An interesting strategy would be to optimise the unitary $\texttt{PREP}(\boldsymbol{\theta})$ directly, circumventing the evaluation of $\mathcal{H}$ and $\mathcal{S}$ altogether. In that case, one evaluates the cost function directly on the QPU, and the matrix elements of the kernels are not accessed directly. Encoding and optimising the linear combination directly reduces the number of expectation values to evaluate from $\mathcal{O}(K^{\alpha=1 \text{ or } 2})$ to only one, at the price of requiring to optimise the vector $\boldsymbol{\theta}$ iteratively. The cost of this approach scales as $(L+K)\cost{layer}\times n_{\text{evals}} / p_0$, where $n_{\text{evals}}$ is the total number of evaluations of the cost function $\mel{\Psi(\boldsymbol{\theta})}{\hat{H}_C}{\Psi(\boldsymbol{\theta})}$ during the variational optimisation, and $p_0$ is the success probability of the encoding. It is therefore a more economic alternative to the Hadamard plus LCU method if $n_{\text{evals}}<4(K-1)(L+1)/(K+L)$. If the kernels do not have a Toeplitz structure, the pure LCU approach becomes favourable if $n_{\text{evals}}<2K^2(L+1)/(K+L)$. Naturally, the number of function evaluations critically depends upon the optimisation algorithm. A general observation is that even for moderate values of $K$, a few tens to hundred of evaluations will typically be needed\footnote{Due to the approximate linear dependence between the trial states, $n_\text{evals}$ may even take larger values.}, making the variational optimisation of the LCU angles unlikely to compete with the Hadamard plus LCU approach for this specific choice of generating operator. Finally, the LCU method naturally lends itself to the determination of excited states using the QSD, with a total complexity quadratic in the number of desired eigenstates.

\section{Numerical results}\label{sec:gcm_results}

\subsection{Parameter setting}
For all the numerical applications shown below, the number of QAOA layers is set to $L=20$, and only one layer is optimised at a time in order to keep the runtime reasonable. That is, once the $\ell^\text{th}$ layer's parameters $(\gamma_\ell,\beta_\ell)$ have been optimised, they are fixed before moving on to the optimisation of the $(\ell+1)^\text{th}$ layer. The angles are optimised using Scipy's implementation of the limited-memory Broyden-Fletcher-Goldfarb-Shanno (L-BFGS-B) gradient descent algorithm~\cite{Byrd1995_Broyden,Zhu_1997_Broyden}. Finally, after the $L$ layers, the optimal depth $L'\leq L$ is chosen simply by selecting the depth that leads to the lowest cost function.

For the QSE stage, the evolution times are taken equally spaced in the $[-\pi(1-1/K);\pi(1-1/K)]$ interval. This gives both $\mathcal{H}$ and $\mathcal{S}$ a Toeplitz structure, such that the number of matrix element to estimate is linear with in $K$. When diagonalising the overlap kernel, states with eigenvalue below $\ecut=10^{-3}$ are discarded. These parameters are chosen somewhat arbitrarily, and several problem-specific improvements are possible. As the QSE is guaranteed to deliver improved solution with respect to the QAOA, the simple choice for the values of the generator coordinates nonetheless leads to substantially better solutions.

\subsection{Preliminary analysis: \texorpdfstring{$G3$}{G3} and \texorpdfstring{$K^+_{3,3}$}{K33} graphs}
\begin{figure}
    \centering
    \includegraphics[width=\linewidth]{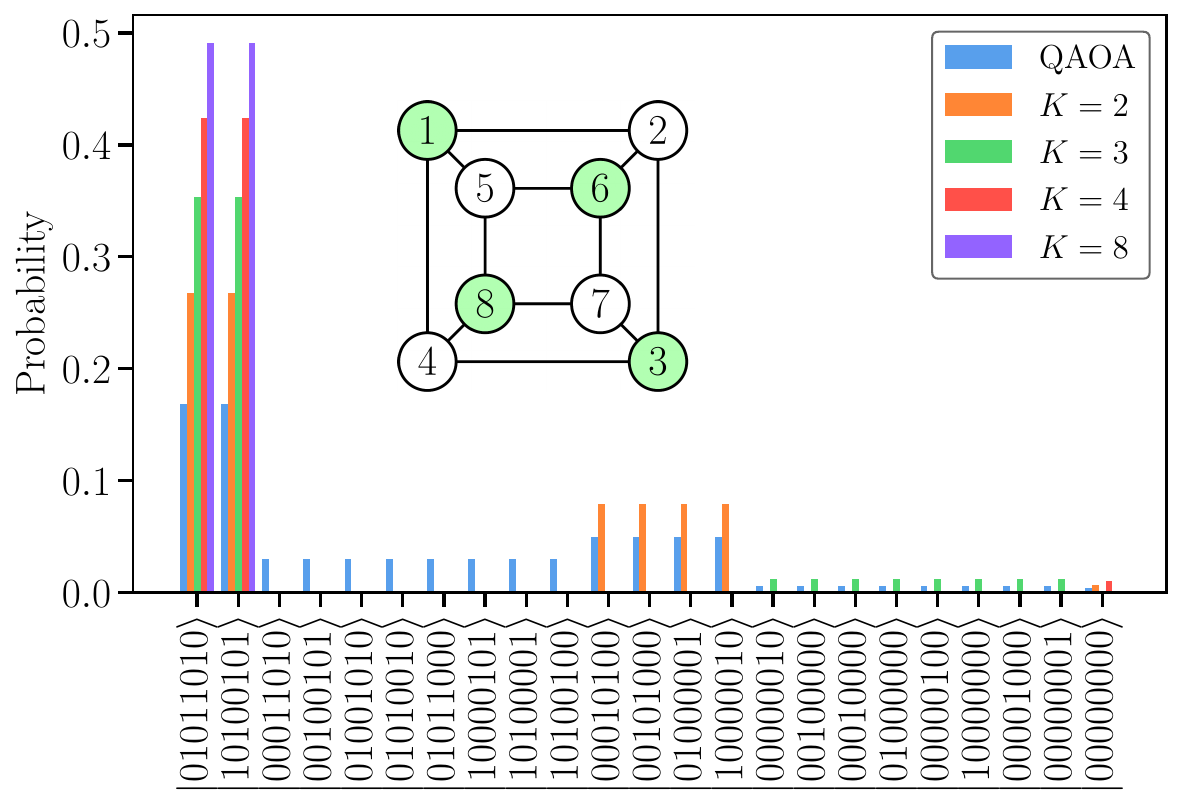}
    \caption{Probabilities of measuring different bit-strings with the QAOA algorithm and with the QSE, for varying number of generating states, on the $G3$ graph. The inset shows the graph, with the nodes coloured in green corresponding to one of the maximally independent sets ($\ket{10100101}$); the other MIS is the complement of the one shown ($\ket{01011010}$). }\label{fig:cube_graph_probabilities}
\end{figure}

\begin{figure}
    \centering
    \includegraphics[width=\linewidth]{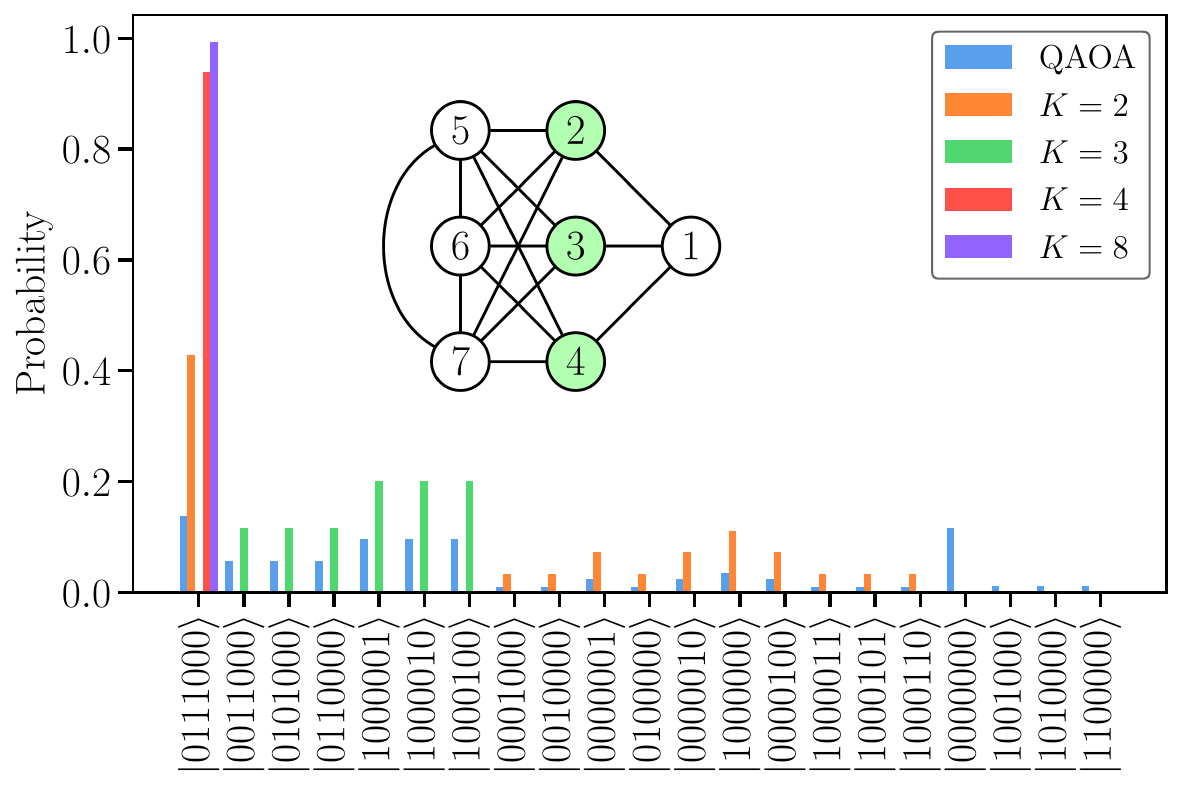}
    \caption{Probabilities of measuring different bit-strings with the QAOA algorithm and with the QSE, for varying number of generating states, on the $K^+_{3,3}$ graph. On the inset, the nodes coloured in green corresponds to the MIS, $\ket{0111000}$.}
    \label{fig:K33p_graph_probabilities}
\end{figure}

In order to analyse in full detail the performances of the QSE over the plain QAOA, I focus on two specific graphs. The first is the cube graph $G3$ (Fig.~\ref{fig:cube_graph_probabilities}), that is, the graph formed by the vertices and edges of a standard die. Do to its symmetries, it is particularly easy to identify the two MIS for this graph. The second graph is $K^+_{3,3}$ (Fig.~\ref{fig:K33p_graph_probabilities}), and has the attractive feature of being the smallest graph for which the best greedy algorithm fails~\cite{dalyac_exploring_2023}. It is therefore expected to be more challenging than the cube graph, and the two make up a minimalistic yet meaningful testbed to asses the efficiency of the QSE.

For both graphs, the QAOA state already strongly overlaps with the lowest energy bit-strings, with approximately $17\%$ for each MIS of the $G3$ graph, and $14\%$ for the $K^+_{3,3}$ graph. The QSE step is found to systematically increase the fidelity for the $G3$ graph, reaching above $98\%$ using $K=8$ generator states. For larger $K$, the fidelity is essentially equal to one. The situation is similar for the more complex $K^+_{3,3}$ graph, with the exception of the $K=3$ case, where the QSE incorrectly cancels out the odd-parity states. At larger $K$, the ground state correctly identified, with a fidelity exceeding $98\%$ at $K=8$.

Besides the increase of the fidelity, another figure of merit of the QSE is that increasing $K$ systematically shifts the amplitudes towards low-energy states, resulting in a lower value of the cost function even when the resulting configuration does not belong in the right symmetry sector (as for the $K^+_{3,3}$ graph with $K=3$). While increasing $K$ guarantees that the ground state(s) will eventually be identified correctly, projecting the QAOA state in either the positive or negative parity subspace prior to switching on the QSE will also magnify the amplitude of the ground state.

\subsection{Erd\"os-Rényi graphs}
To showcase and benchmark the capabilities of the QSE more systematically, I apply the machinery on Erd\"os-Rényi (ER) graphs of average density $\rho=0.5$. These graphs are NP-hard in general~\cite{dalyac_exploring_2023}. In particular, they cannot be addressed using analogue quantum processors when the number of nodes becomes large, as the probability that a random ER graph with fixed non-zero density is a unit-ball graph approaches zero in the large $N$ limit~\cite{cazals_identifying_2025}. Although for small graph size, the probability of generating a disconnected graph is non-negligible, this probability decreases exponentially fast as $N$ grows and $\rho$ remains fixed and strictly nonzero. For ER graphs, there is therefore no practical need to identify whether a graph is connected or not~\cite{kim_quantum_2024,cazals_identifying_2025}. 

\begin{figure*}[t!]
    \centering
    \includegraphics[width=\textwidth]{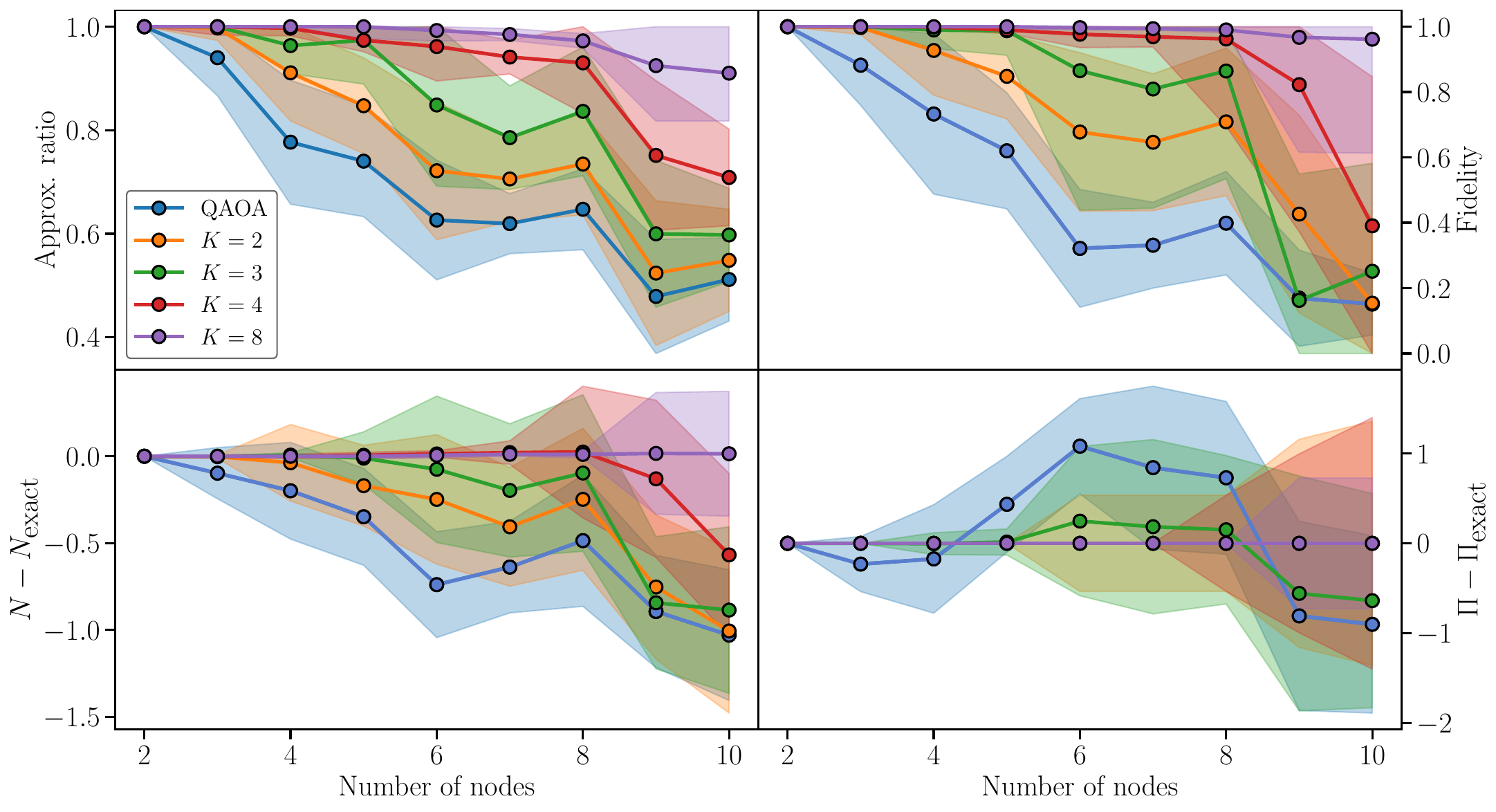}
    \caption{Approximation ratio (top left), fidelity (top right), error in the Hamming weight/particle number (bottom left), and error in the parity (bottom right). For each number of nodes, 14 random Erd\"os-Rényi graphs are generated. The shaded areas correpond to one standard deviation.}
    \label{fig:gcm_results}
\end{figure*}

The QAOA-plus-QSE workflow is applied to graphs of size $N=2$ to $10$. In addition to the approximation ratio and the fidelity, the error in the particle number (i.e. in the Hamming weight) and parity are also shown on Fig.~\ref{fig:gcm_results}.

The QAOA achieves rather poor approximation ratios, reaching $0.5$ for $N=10$. Naturally, the fidelity degreades to even lower values, namely $0.15$ at $N=10$. Furthermore, the error in both the Hamming weight and the parity of the QAOA solutions is of order one, indicating that QAOA fails to evolve the initial state towards the correct symmetry subspace. 

Adding the QSE step resolves all the issues of the QAOA optimisation, and increasing the number of trial states systematically improves all the relevant performance metrics. Notably, for $N=10$, the approximation ratio is almost double. The performance of the QSE is even more striking for the fidelity, which remains significantly higher that the QAOA solutions, with a value of $0.96$ for $N=10$ nodes. As a consequence of the high fidelity, the Hamming weight (and therefore the parity) of the QSE solution quickly converge to the exact values.

Fitting the fidelities of the QAOA and the QSE to Fermi-Dirac distributions, one can extract the critical graph size $N_*$ beyond which the QSE presents a favourable time-to-solution as the smallest $N$ for which
\begin{align}
    \frac{\beta_\text{QAOA}}{1+e^{N\alpha_\text{QAOA}}}\frac{1+e^{N\alpha_\text{QSE}(K)}}{\beta_\text{QSE}(K)} > 2K(\sqrt{\rho}N)^3,
\end{align}
assuming the Hadamard plus Pauli method is used. In that specific setting and for $K=8$, the QAOA-plus-QSE surpasses QAOA for $N > 75$, and increasing $K$ should further reduce $N_*$. It should be stressed that this number implicitly depends on the choice of the QAOA initial state and mixer, on the number of layers, on the optimisation algorithm and strategy, on the type of graphs, and on the optimisation problem itself. As such, the precise value obtained here bears little meaning, and the only merit of such extrapolation is to illustrate the advantages of the QSE over QAOA for graphs with size of practical interest.

\section{Conclusion}

In this work, the generator coordinate method is applied to the maximal independent set problem of graph theory. The initial state of the GCM is obtained using the QAOA. The QAOA-plus-GCM workflow is applied to two illustrative graphs, and to Erd\"os-Rényi graphs of varying size and constant average density. It is shown to systematically yield improved solutions compared with the plain QAOA, and in particular, the cost-to-solution of the proposed method is estimated to cross the QAOA's for the critical graph size $N_*=75$, beyond which the algorithm outperforms QAOA. Analysing the scaling and formal properties of the GCM, $N_*$ can be expected to remain within the same order of magnitude regardless of the specificities of the cost and mixer Hamiltonians, and can therefore be anticipated to find purpose for several kinds of combinatorial optimisation problems. Furthermore, as the GCM approach is essentially agnostic to which initial state preparation algorithm is employed, it is likely to find promising applications as an additional step on top of several other algorithms and frameworks, such as analogue computing. In particular, GCM could be employed to speed up adiabatic evolution while offering convergence guarantees due to its purification properties. Altogether, the versatility and performance guarantees of the GCM  make it an attractive algorithm for combinatorial optimisation problems.

\section{Acknowledgements}

I am grateful to Denis Lacroix and Jing Zhang for useful discussions on an earlier version of this article. In particular, D.L. for suggested the possibility to optimise the LCU angles iteratively and kindly proofread the final paper. I also wish to warmly thank E. Beaujeault-Taudière for his enthusiastic comments at all stages of this work.

\appendix

\section{Comparison of the different kernel estimation algorithms}\label{app:gate_counts}

\subsection{Cost of controlled coherent rotations}
\subsubsection{Decomposition of the circuits}
The cost of each of the three methods  breaks down in three parts, namely the cost of the QAOA state preparation, the cost of the circuit itself, and a final multiplicative factor accounting for either the success probability or the increase in the number of shots, that compensates the increase in the variance when summing different expectation values. With this splitting, one has, at first order in $N$,
\begin{widetext}
\begin{align}
    &\left[\cost{QAOA}+\frac{\rho N^2}{2} \cost{$CR_{ZZ}$}\right] \times (\sqrt{\rho}N)^3 \quad \text{(Pauli)}, \label{eq:circuits_costs_Pauli}\\
    &\left[\cost{QAOA}+\frac{\rho N^2}{2}\cost{$CR_{ZZ}$}\right] \times p C(p)\quad \text{(RTE)}, \label{eq:circuits_costs_RTE}\\
    &\left[\cost{QAOA}+\rho N^2\cost{$CZZ$}\right] \times \frac{\rho^2 N^4}{4\expval{\hat{H}^2_C}}\quad \text{(LCU)}.\label{eq:circuits_costs_LCU}
\end{align}
\end{widetext}

The encoding of the QAOA solution has cost
\begin{align}
    \cost{QAOA}=\frac{\rho N^2}{2}\cost{$R_{ZZ}$} \times L',
\end{align}
where $L' \leq L$ is the number of layers retained after the optimisation.

\begin{figure}
    \centering
    \resizebox{0.4\linewidth}{!}{\input{circuits_sketches/LCU_ctrls_grouping/without_grouping}}\\
    \vspace{10mm}
    \resizebox{\linewidth}{!}{\input{circuits_sketches/LCU_ctrls_grouping/without_grouping_decomposed}}\\

    \caption{Top: circuit encoding the two operators $V_k$ and $V_{k+1}$ on a register. The half-filled dots (\halfctrl) represent a set of controls whose value is irrelevant to the working of the circuit. Bottom: decomposition of the circuit into a sequence of multi-controlled NOT gates and single-controlled unitaries $CV_k$ and $CV_{k+1}$. The two central controls enclosed in the braces cancel each other and can thus be removed.}
    \label{fig:LCU_controls_grouping}
\end{figure}
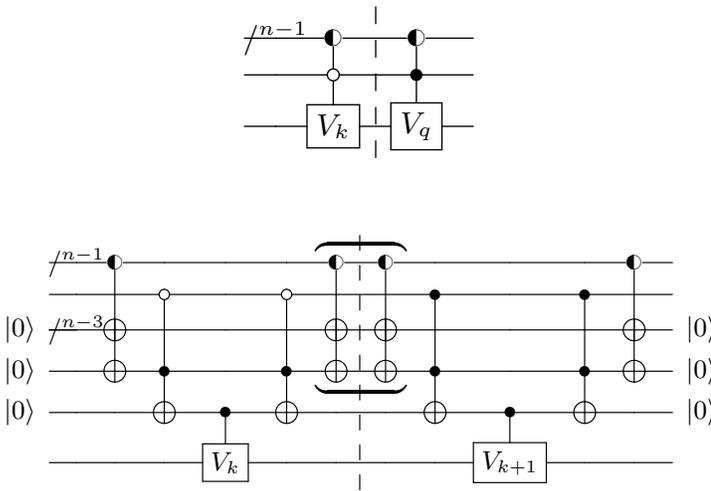

For the RTE, the parameter $1/2\leq C(p) \leq \pi^2/3$ is the sum of the squared finite difference coefficients~\cite{wiki:Finite_difference_coefficient}. For the LCU circuit, the cost is not proportional to $N^2$Cost$(C^nZZ)$ as would be given by simply adding the cost of each $C^nZZ$, but rather to $N^2$Cost$(CZZ)$ because the $\{V_k\}$ operators can be rearranged such that the controls of $V_k$ and $V_{k+1}$ differ only by the value of one bit, as shown on Fig.~\ref{fig:LCU_controls_grouping}. Thus, only two Toffoli gates must be applied between the controlled $V_k$ and the controlled $V_{k+1}$, instead of the $2(n-1)$ naively required.

\subsubsection{Cost of \texorpdfstring{$R_{ZZ}, CR_{ZZ}$}{RZZ, CRZZ} and \texorpdfstring{$C^nZZ$}{CnZZ}}
These three unitaries can be decomposed in several ways, and which one is optimal depends upon several factors. Naturally, the first one is the elementary gate set the QPU; I focus on the Clifford+$T$ gates, which is the set of primitives targeted by most of the FTQC architectures.

For the $R_{ZZ}$ and $CR_{ZZ}$ gates, the second key element is the precision to which the rotations are implemented and whether their decomposition uses ancillas~\cite{ross2016optimalancillafreecliffordtapproximation}. Here, I consider the optimal ancilla-free algorithm of ~\cite{ross2016optimalancillafreecliffordtapproximation}, which allows decomposing any one-qubit rotation with error smaller than $\varepsilon$ using $\approx4\log(1/\varepsilon)$ $T$ gates.

\begin{figure}[ht!]
    \begin{center}
        \resizebox{\linewidth}{!}{\input{circuits_sketches/CnRZZ_and_decomposition}}\\
        \vspace{10pt}
        \resizebox{\linewidth}{!}{\input{circuits_sketches/CRZZ}}
    \end{center}
    \caption{Top: decomposition of a $C^nR_{ZZ}$ gate into a sequence of $C^nX$ gates and a central $CR_{ZZ}$ gate~\cite{Nielsen_Chuang10}. Bottom: decomposition of a $CR_{ZZ}(\theta)$ gate into a sequence of CNOT and single-qubit rotations.}
    \label{fig:multictrl_and_CRZZ}
\end{figure}
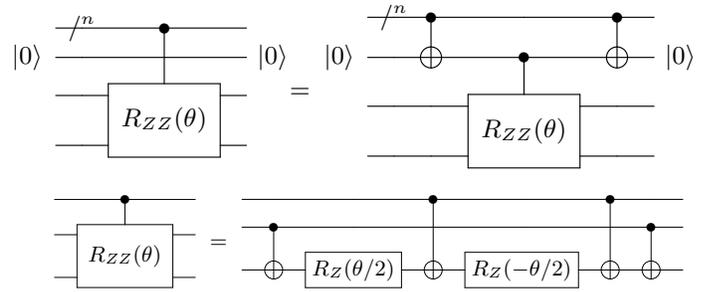

Regarding $C^nZZ$, an efficient~\cite{Barenco_1995,Nielsen_Chuang10} implementation of the $C^nZZ$ gate is to decompose it into a sequence of two $C^nX$ surrounding a central $CZZ$ gate, as shown in Fig.~\ref{fig:multictrl_and_CRZZ}. 
The $C^nX$ gates can then be implemented in multiple manners~\cite{Barenco_1995,Nielsen_Chuang10,Gidney_2015_Constructing,Claudon_2024}. Which decomposition is optimal will, among others, be dictated by hardware-related considerations, lying outside the scope of the present work. The final gate count given in table~\ref{tab:gate_counts} reflects a specific choice, namely, decoposing each $C^nX$ into an array of $(n-1)$ Toffoli gates using zeroed ancillas. Eventually, each Toffoli gate is further decomposed into 6 CNOT and 7 $T$ gates~\cite{Nielsen_Chuang10}, leading to the final cost of $12(n-1)$ CNOT and $14(n-1)$ $T$ gates.

The gate counts for these three elementary operations are given in Table~\ref{tab:basic_gate_counts}. Combining these with \eqref{eq:circuits_costs_RTE}--\eqref{eq:circuits_costs_Pauli} yields the leading order CNOT and $T$ gates costs of the three different methods, as summarised in Table~\ref{tab:gate_counts}.
\begin{table*}[t!]
    \centering
    \begin{tabular}{|c|c|c|c|c|c}
        \hline
        \hline
        Gate & CNOT & $T$ & Toffoli & Ancillas\\
        \hline
        $R_{ZZ}$ & 2 & $8\log_2(1/\varepsilon)$ & 0 & 0\\
        $CR_{ZZ}$ & 4 & $8\log_2(1/\varepsilon)$ & 0 & 0\\
        $C^n{ZZ}$ & $12(n-1)+4$ & $14(n-1)$ & $2(n-1)$ & $n-1$ \\
        \hline
        \hline
    \end{tabular}
    \caption{Decomposition of the three basic gates into CNOT and $T$ gates. The number of controls $n$ is approximately equal to $\lceil 2\log_2(N)\rceil$. The numbers of Toffoli gates and ancillas are also given for completeness.}
    \label{tab:basic_gate_counts}
\end{table*}

\begin{table*}[t!]
    \centering
    \begin{tabular}{|c|c|c|c|c|}
        \hline 
        \hline 
        Method & CNOT & $T$ & Toffoli & Ancillas\\
        \hline 
        Pauli & $(\sqrt{\rho} N)^5 (L'+2) $ & $4 (\sqrt{\rho} N)^5 (L'+1) \log_2(1/\varepsilon)$ & 0 & 1 \\
        \hline
        RTE  & $\rho N^2 (L'+2) p C(p)$ & $(4\rho N^2 (L'+1) \log_2(1/\varepsilon)) p C(p)$ & 0 & 1 \\
        \hline 
        LCU & $\rho^3 N^6  (L' + 4) / (4\expval{\hat{H}^2_C})$ &  $\rho^3 N^6 L'\log_2(1/\varepsilon) / \expval{\hat{H}^2_C}$ & $4\log_2(N)$ & $4\log_2(N)$ \\
        \hline
        \hline 
    \end{tabular}
    \caption{Gate counts of the different methods for the evaluation of a single expectation value. The numbers are given to leading order in $N$.
    }
    \label{tab:gate_counts}
\end{table*}

\subsection{Comparison of the methods}
A first observation is that comparing the RTE and the Pauli circuits is straightforward; which one requires fewer resources is simply given by whether $pC(p)/(\sqrt{\rho}N)^3$ is smaller or greater than one. However, which value of $p$ is appropriate is problem-dependent, and rather delicate to estimate beforehand due to the interplay between $t,p,\langle H^p \rangle$ and the number of measurements. A first remark is that the precision cannot be increased arbitrarily by reducing $t$, since the total number of measurements, $N_\text{shots}$, should be larger than $p/t^2$ in order to resolve the matrix elements accurately. Even with $t$ fixed, the difficulty of estimating the off-diagonal moments $\expval{\hat{H}_C^p}$ makes the selection of $p$ difficult. It can be noted, however, that the scaled approximation ratio of finite-depth QAOA decreases with the graph size, indicating that the value of the moments will concentrate to a multiple the spectral radius of $\hat{H}_C$, which grows quadratically with $N$. This in turn implies that for the finite differences error to decrease with $p$, the number of measurements should scale as $\rho^2 N^4$. Another estimate is obtained by taking the average cost over all bitstrings, which also grows quadratically. This gives the scaling $p \propto N^4 / N_\text{shots}$. If $N_\text{shots}$ is sufficiently large however, such that $p \sim 1$, then the lower bound $\expval{\hat{H}_C^p} \approx \expval{\hat{H}_C}^p$ can be used to select $p$. In the case of the LCU, similarly, $\expval{\hat{H}_C^2} \geq \expval{\hat{H}_C}^2$ gives an upper bound to estimate the cost of the algorithm.

For the MIS problem specifically, an optimistic assummption is to expect that the QAOA is able to find a solution close to the optimum, such that $\expval{\hat{H}_C^p}\propto \log^p(N)$. In that case, the RTE, LCU, and Pauli methods respectively have gate counts scaling as $N^2 \log^p(N), N^6/\log^2(N)$, and $N^5$, and the RTE is therefore the most efficient for large $N$. As a side remark, for quantum many-body problems, one generally achieves $\expval{\hat{H}_C}\propto N$ (for instance using mean-field theory), and the Hadamard plus LCU methods then offers the best scaling.

\section{Short note about the truncation method}\label{app:gcm_reencoding}

The Hill-Wheeler (HW) equation
\begin{align}
    \mathcal{H}\ket{\Psi}=E\mathcal{S}\ket{\Psi}
\end{align}
is often ill-conditioned due to the trial vectors being nearly linearly dependent. The ``truncation method''~\cite{bonche_analysis_1990,epperly_theory_2023} is a simple procedure that defines a smaler orthonormal basis from said vectors. It consists in diagonalising the overlap kernel as $\mathcal{S}=P\Sigma P^\dag$, removing the columns of $P$ associated to eigenvalues below some threshold $\ecut$ (yieding the truncated transformation $\tilde{P}$), and projecting the HW equation into that new basis\footnote{To be rigorous, all quantities in that equation should be written with a tilde, since they are not the same as in the original equation. But I omit this for convenience.}:
\begin{align}
    \tilde{\mathcal{H}}\ket{\tilde{\Psi}}&=E\tilde{\mathcal{S}}\ket{\tilde{\Psi}},
\end{align}
with $\tilde{\mathcal{H}} \equiv P \mathcal{H} P^\dag$  and $\tilde{\mathcal{S}} \equiv P \mathcal{S} P^\dag$. The eigenstates can be reconstructed in two steps:
\begin{align}
    \ket{\Psi} &= \sum_j a_j \ket{\tilde{\Phi}_j} = \sum_j a_j \sum_{k,j} \tilde{P}_{jk} \ket{\chi_k}\nonumber\\
    &= \sum_{k,j} \left(\sum_j a_j \tilde{P}_{jk}\right) \ket{\chi_k}.
\end{align}

\section{Kernels-free GCM}\label{app:kernels_free_gcm}

An alternative approach that does requires neither the evaluation of the kernels~\eqref{eq:kernels_definition_H}\eqref{eq:kernels_definition_S} nor the variational optimisation of the weights is to select the $\{f_{k,j}\}$ such that the configuration mixing implements a Gaussian filter: in the case of two trial states, and after shifting the cost Hamiltonian according to $\hat{H}_C \rightarrow \hat{H}_C - E_j$,
\begin{align}
    V(t^2) &\propto \frac{U(t) + U(-t)}{2}\nonumber\\
    &= e^{-t^2 (\hat{H}_C-E_j)^2} + \mathcal{O}((\hat{H}_C-E_j)^4 t^4),
\end{align}
where $U(t)=\exp(-i(\hat{H}_C-E_j)t)$. Iterating the linear combination gives, for $K$ trial states with evenly spaced coordinates,
\begin{align}\label{eq:qite}
    V(Kt^2) \propto \frac{1}{2^{K}}\sum^{K-1}_{k=0} C^k_K U((K-1-2k)t).
\end{align}
The right-hand side of \eqref{eq:qite} can be implemented using the LCU circuit of Fig.~\ref{fig:LCU} with probability
\begin{align}
    p^{(K)}_0 &= \mel{\Psi_0}{e^{-2Kt^2(\hat{H}_C-E_j)^2}}{\Psi_0} \nonumber\\
    &+ \mathcal{O}(K^2(\hat{H_C}-E_j)^4t^4).
\end{align}
Provided the shift is suitably chosen, this probability can be made close to one. Likewise, starting from
\begin{align}
    W(t) &\propto \frac{(1+i)U(-t)+(1-i)U(t)}{2} \nonumber\\
    &= e^{-(\hat{H}_C-E_j)t} + \mathcal{O}((\hat{H}_C-E_j)^2t^2),
\end{align}
it is possible to realise an imaginary time evolution
\begin{align}
    &W(Kt) \approx W^K(t) \nonumber\\
    &\propto \sum^{K-1}_{k=0} C^k_K \left(\frac{1-i}{2}\right)^k \left(\frac{1+i}{2}\right)^{K-1-k} U((K-1-2k)t).
\end{align}
The value of the shift determines the centre of the Gaussian, and should not be too far from the smallest eigenvalue of $\hat{H}_C$. It can for instance be estimated from the QAOA energy or the asymptotic scaling of the MIS size. Furthermore, it is possible to compute new guesses of the shift for increasing values of $K$, which ensures $p^{(K)}_0$ remains of order one even for large $K$.

A side effect of this construction is that it shows that configuration mixing \eqref{eq:gcm_expansion} with real-time-evolved states improves the fidelity exponentially quickly with the number of trial states.

\bibliography{bibliography.bib}

\end{document}

%% file: circuits_sketches/hadamard_test.tex
\Qcircuit @C=1em @R=.7em { 
\lstick{\ket{0}}  & \gate{H} & \gate{R_z(\phi)} & \ctrl{1} & \ctrl{1} & \ctrl{1} & \gate{H} & \meter\\ 
\lstick{\ket{\Psi_0}} & \qw{/}^{N} & \qw & \gate{e^{-i\hat{H}_C t_{k'}}} & \gate{\hat{A}} & \gate{e^{+i\hat{H}_C t_k}} & \qw & \qw \\
}

%% file: circuits_sketches/hadamard_Pauli.tex
$
\Qcircuit @C=1em @R=.7em { 
\lstick{\ket{0}}  & \gate{H} & \gate{R_z(\phi)} & \ctrl{1} & \ctrl{1} & \gate{H} & \meter\\ 
\lstick{\ket{\Psi_0}} & \qw{/}^{N} & \qw & \gate{e^{-i\hat{H}_C (t_{k'}-t_k)}} & \gate{\sum_{(n,n')\in E}Z_nZ_{n'}}& \qw & \qw\\
}
$

%% file: circuits_sketches/hadamard_RTE.tex
$
\Qcircuit @C=1em @R=.7em { 
\lstick{\ket{0}}  & \gate{H} & \gate{R_z(\phi)} & \ctrl{1} & \gate{H} & \meter\\ 
\lstick{\ket{\Psi_0}} & \qw{/}^{N} & \qw & \gate{e^{-i\hat{H}_C (t_j+t_{k'}-t_{k})}} & \qw & \qw\\
}
$

%% file: circuits_sketches/Hadamard_LCU.tex
\newcommand{\up}[1]{\push{\raisebox{6pt}{$#1$}}}
\newcommand{\prep}{\text{PREP}(\boldsymbol{\theta})}
\newcommand{\prepdag}{\text{PREP}^\dag(\boldsymbol{\theta})}
\newcommand{\sel}{\text{SEL}}

$
\Qcircuit @C=0.7em @R=0.7em {
\lstick{\ket{0}} & \qw & \gate{H} & \ctrl{1} & \qw & \qw\barrier[0.2em]{2} & \qw & \ctrl{2} & \gate{H} & \meter\\
\lstick{\ket{0}^{\otimes \lceil \log_2(M)\rceil}} & \qw & \gate{\prep} & \multigate{1}\sel & \gate{\prepdag} & \meter \\
\lstick{\ket{\Phi_0}} & \qw{/}^{N} & \qw & \ghost\sel & \qw & \qw & \qw & \gate{e^{-iH(t_{k'}-t_k)}}
}
$

%% file: circuits_sketches/LCUv2.tex
\newcommand{\up}[1]{\push{\raisebox{6pt}{$#1$}}}
\newcommand{\prep}{\text{PREP}(\boldsymbol{\theta})}
\newcommand{\prepdag}{\text{PREP}^\dag(\boldsymbol{\theta})}

\Qcircuit @C=0.7em @R=0.7em {
\lstick{\ket{0}} & \multigate{3}\prep & \ctrlo{1} & \ctrlo{1} & \qw & \cdots & & \ctrl{1} & \multigate{3}\prepdag & \meter \\
\lstick{\ket{0}} & \ghost\prep & \ctrlo{1} & \ctrlo{1} & \qw & \cdots & & \ctrl{1} & \ghost\prepdag & \meter \\
\lstick{\raisebox{6pt}{\vdots}~\,} & \pureghost\prep & \up{\vdots} & \up{\vdots} & & \up{\ddots} & & \up{\vdots} & \pureghost\prepdag & \raisebox{6pt}{\vdots} \\
\lstick{\ket{0}} & \ghost\prep & \ctrlo{-1} \qwx[1] & \ctrl{-1} \qwx[1] & \qw & \cdots & & \ctrl{-1} \qwx[1] & \ghost\prepdag & \meter \\
\lstick{\ket{\Phi_0}} & \qw{/}^{N} & \gate{V_0} & \gate{V_1} & \qw & \cdots & & \gate{V_{K-1}} & \qw & \qw
}

%% file: circuits_sketches/LCU_ctrls_grouping/without_grouping.tex



\Qcircuit @C=1em @R=.7em {
  \lstick{} & \qw{/}^{n-1} & \push{\halfctrl}\ctrlo{1} \barrier[-1.3em]{2} & \push{\halfctrl}\ctrlo{1} & \qw & \\
  \lstick{} & \qw & \ctrlo{1} & \ctrl{1} & \qw \\
  \lstick{} & \qw & \gate{V_k} & \gate{V_{q}} & \qw
}

%% file: circuits_sketches/LCU_ctrls_grouping/without_grouping_decomposed.tex



\Qcircuit @C=1em @R=.7em {
  \lstick{} & \qw{/}^{n-1} & \push{\halfctrl}\ctrlo{3} & \qw & \qw & \qw & \push{\halfctrl}\ctrlo{3}\barrier{5} & \push{\halfctrl}\ctrlo{3} & \qw & \qw & \qw & \push{\halfctrl}\ctrlo{3} & \qw \\
  \lstick{} & \qw & \qw & \ctrlo{2} & \qw & \ctrlo{2} & \qw & \qw & \ctrl{2} & \qw & \ctrl{2} & \qw & \qw \\
  \lstick{\ket{0}} & \qw{/}^{n-3}\qw & \targ & \qw & \qw & \qw & \targ & \targ & \qw & \qw & \qw & \targ & \qw & \rstick{\hspace{-1em}\ket{0}} \\
  \lstick{\ket{0}} & \qw & \targ & \ctrl{1} & \qw & \ctrl{1} & \targ & \targ & \ctrl{1} &\qw & \ctrl{1} & \targ & \qw & \rstick{\hspace{-1em}\ket{0}} \\
  \lstick{\ket{0}} & \qw & \qw & \targ & \ctrl{1} & \targ & \qw & \qw & \targ & \ctrl{1} & \targ & \qw & \qw & \rstick{\hspace{-1em}\ket{0}} \\
  & \qw & \qw & \qw & \gate{V_{k}} & \qw & \qw & \qw & \qw &  \gate{V_{k+1}} & \qw & \qw & \qw
  \gategroup{1}{7}{4}{8}{.7em}{_)}\gategroup{1}{7}{4}{8}{.7em}{^)}
}

%% file: circuits_sketches/CnRZZ_and_decomposition.tex
$
\begin{array}{c c c}
    \vcenter{
        \input{circuits_sketches/CnRZZ}
    } & = & \hspace{5mm} \vcenter{
        \input{circuits_sketches/multictrl1_v2}
    } 
\end{array}
$

%% file: circuits_sketches/CRZZ.tex
$
\begin{array}{c c c}
    \vcenter{
        \Qcircuit @C=1em @R=1em { 
        & \ctrl{1} & \qw \\
        & \multigate{1}{R_{ZZ}(\theta)} & \qw \\
        & \ghost{R_{ZZ}(\theta)} & \qw \\
        }
    } & = & \vcenter{
        \Qcircuit @C=1em @R=1em { 
        & \qw & \qw & \ctrl{2} & \qw & \ctrl{2} & \qw & \qw \\
        & \ctrl{1} & \qw & \qw & \qw & \qw & \ctrl{1} & \qw \\
        & \targ & \gate{R_Z(\theta/2)} & \targ & \gate{R_Z(-\theta/2)} & \targ & \targ & \qw \\
        }
    }
\end{array}
$

%% file: main.bbl
\begin{thebibliography}{10}

\bibitem{bhattacharya2005graph}
S.~Bhattacharya, A.~Raghunathan, and S.~Singh, ``A graph-theoretic approach to protein structure prediction,'' {\em arXiv preprint q-bio/0510033}, 2005.

\bibitem{mendes2005protein}
P.~Mendes, M.~Nunes, and S.~Oliveira, {\em Protein Interaction Networks and Computational Biology}.
\newblock Springer, 2005.

\bibitem{aluru2006handbook}
S.~Aluru, {\em Handbook of Computational Molecular Biology (1st ed.)}.
\newblock Chapman and Hall/CRC, 2005.

\bibitem{stauffer1994percolation}
D.~Stauffer and A.~Aharony, {\em Introduction to Percolation Theory}.
\newblock Taylor \& Francis, 1994.

\bibitem{weigt2001hardspherefluidsolid}
M.~Weigt and A.~K. Hartmann, ``Minimal vertex covers on finite-connectivity random graphs: A hard-sphere lattice-gas picture,'' {\em Physical Review E}, vol.~63, Apr. 2001.

\bibitem{korte2006combinatorial}
B.~Korte and J.~Vygen, {\em Combinatorial Optimization: Theory and Algorithms}.
\newblock Springer, 2006.

\bibitem{hillier2005introduction}
F.~S. Hillier and G.~J. Lieberman, {\em Introduction to Operations Research}.
\newblock McGraw-Hill, 8th~ed., 2005.

\bibitem{button2010handbook}
K.~Button, {\em Handbook of Transportation Science}.
\newblock Springer, 2010.

\bibitem{toth2002vehicle}
P.~Toth and D.~Vigo, ``Vehicle routing problems: A survey,'' {\em Computers \& Operations Research}, vol.~29, no.~3, pp.~187--213, 2002.

\bibitem{nebel2008vlsi}
W.~Nebel and R.~Drechsler, {\em VLSI Design and Test}.
\newblock Springer, 2008.

\bibitem{sherwani1999algorithms}
N.~Sherwani, {\em Algorithms for VLSI Physical Design Automation}.
\newblock Kluwer Academic Publishers, 1999.

\bibitem{johnson_quantum_2011}
M.~W. Johnson, M.~H.~S. Amin, S.~Gildert, T.~Lanting, F.~Hamze, N.~Dickson, R.~Harris, A.~J. Berkley, J.~Johansson, P.~Bunyk, E.~M. Chapple, C.~Enderud, J.~P. Hilton, K.~Karimi, E.~Ladizinsky, N.~Ladizinsky, T.~Oh, I.~Perminov, C.~Rich, M.~C. Thom, E.~Tolkacheva, C.~J.~S. Truncik, S.~Uchaikin, J.~Wang, B.~Wilson, and G.~Rose, ``Quantum annealing with manufactured spins,'' {\em Nature}, vol.~473, pp.~194--198, May 2011.

\bibitem{ebadi_quantum_2022}
S.~Ebadi, A.~Keesling, M.~Cain, T.~T. Wang, H.~Levine, D.~Bluvstein, G.~Semeghini, A.~Omran, J.~Liu, R.~Samajdar, X.-Z. Luo, B.~Nash, X.~Gao, B.~Barak, E.~Farhi, S.~Sachdev, N.~Gemelke, L.~Zhou, S.~Choi, H.~Pichler, S.~Wang, M.~Greiner, V.~Vuletic, and M.~D. Lukin, ``Quantum {Optimization} of {Maximum} {Independent} {Set} using {Rydberg} {Atom} {Arrays},'' {\em Science}, vol.~376, pp.~1209--1215, June 2022.
\newblock arXiv:2202.09372 [quant-ph].

\bibitem{dalyac_exploring_2023}
C.~Dalyac, L.-P. Henry, M.~Kim, J.~Ahn, and L.~Henriet, ``Exploring the impact of graph locality for the resolution of {MIS} with neutral atom devices,'' June 2023.
\newblock arXiv:2306.13373 [quant-ph].

\bibitem{kim_quantum_2024}
K.~Kim, M.~Kim, J.~Park, A.~Byun, and J.~Ahn, ``Quantum {Computing} {Dataset} of {Maximum} {Independent} {Set} {Problem} on {King}'s {Lattice} of over {Hundred} {Rydberg} {Atoms},'' {\em Scientific Data}, vol.~11, p.~111, Jan. 2024.
\newblock arXiv:2311.13803 [quant-ph].

\bibitem{leclerc_implementing_2024}
L.~Leclerc, C.~Dalyac, P.~Bendotti, R.~Griset, J.~Mikael, and L.~Henriet, ``Implementing transferable annealing protocols for combinatorial optimisation on neutral atom quantum processors: a case study on smart-charging of electric vehicles,'' Nov. 2024.
\newblock arXiv:2411.16656 [quant-ph].

\bibitem{Amaro2021}
D.~Amaro, C.~Modica, M.~Rosenkranz, M.~Fiorentini, M.~Benedetti, and M.~Lubasch, ``Filtering variational quantum algorithms for combinatorial optimization,'' {\em arXiv preprint arXiv:2106.10055}, 2021.
\newblock Presents a filtering-enhanced VQE variant on gate-based quantum hardware to solve optimization problems.

\bibitem{Harrigan2021}
M.~P. Harrigan, K.~J. Sung, M.~Neeley, K.~J. Satzinger, F.~Arute, K.~Arya, J.~Atalaya, J.~C. Bardin, R.~Barends, S.~Boixo, M.~Broughton, B.~B. Buckley, D.~A. Buell, B.~Burkett, N.~Bushnell, Y.~Chen, Z.~Chen, B.~Chiaro, R.~Collins, W.~Courtney, S.~Demura, A.~Dunsworth, D.~Eppens, A.~Fowler, B.~Foxen, C.~Gidney, M.~Giustina, R.~Graff, S.~Habegger, A.~Ho, S.~Hong, T.~Huang, L.~B. Ioffe, S.~V. Isakov, E.~Jeffrey, Z.~Jiang, C.~Jones, D.~Kafri, K.~Kechedzhi, J.~Kelly, S.~Kim, P.~V. Klimov, A.~N. Korotkov, F.~Kostritsa, D.~Landhuis, P.~Laptev, M.~Lindmark, M.~Leib, O.~Martin, J.~M. Martinis, J.~R. McClean, M.~McEwen, A.~Megrant, X.~Mi, M.~Mohseni, W.~Mruczkiewicz, J.~Mutus, O.~Naaman, C.~Neill, F.~Neukart, M.~Y. Niu, T.~E. O'Brien, B.~O'Gorman, E.~Ostby, A.~Petukhov, H.~Putterman, C.~Quintana, P.~Roushan, N.~C. Rubin, D.~Sank, A.~Skolik, V.~Smelyanskiy, D.~Strain, M.~Streif, M.~Szalay, A.~Vainsencher, T.~White, Z.~J. Yao, P.~Yeh, A.~Zalcman, L.~Zhou, H.~Neven, D.~Bacon, E.~Lucero, E.~Farhi, and R.~Babbush, ``Quantum
  approximate optimization of non-planar graph problems on a planar superconducting processor,'' {\em Nature Physics}, vol.~17, pp.~332--336, Mar 2021.

\bibitem{Jain2022}
N.~Jain, B.~Coyle, E.~Kashefi, and N.~Kumar, ``Graph neural network initialization of quantum approximate optimization,'' {\em Quantum}, vol.~6, 2022.
\newblock Uses a graph neural network to initialize QAOA circuits for improved performance on digital quantum computers.

\bibitem{Zhou2022}
Z.~Zhou, Y.~Du, X.~Tian, and D.~Tao, ``Qaoa-in-qaoa: Solving large-scale maxcut problems on small quantum machines,'' {\em arXiv preprint arXiv:2205.11762}, 2022.
\newblock Proposes a QAOA decomposition framework to solve large MaxCut problems on limited-qubit gate-based quantum computers.

\bibitem{Ponce2023}
M.~Ponce, R.~Herrman, P.~C. Lotshaw, S.~Powers, G.~Siopsis, T.~Humble, and J.~Ostrowski, ``Graph decomposition techniques for solving combinatorial optimization problems with variational quantum algorithms,'' {\em arXiv preprint arXiv:2306.00494}, 2023.
\newblock Applies graph decomposition strategies to make combinatorial problems tractable for QAOA on gate-based quantum computers.

\bibitem{Dupont2023}
M.~Dupont, B.~Evert, M.~J. Hodson, B.~Sundar, S.~Jeffrey, Y.~Yamaguchi, D.~Feng, F.~B. Maciejewski, S.~Hadfield, M.~S. Alam, Z.~Wang, S.~Grabbe, P.~A. Lott, E.~G. Rieffel, D.~Venturelli, and M.~J. Reagor, ``Quantum-enhanced greedy combinatorial optimization solver,'' {\em Science Advances}, vol.~9, no.~45, 2023.
\newblock Implements a hybrid quantum-classical greedy optimization algorithm on a superconducting gate-based quantum processor.

\bibitem{farhi_quantum_2014}
E.~Farhi, J.~Goldstone, and S.~Gutmann, ``A {Quantum} {Approximate} {Optimization} {Algorithm},'' Nov. 2014.
\newblock arXiv:1411.4028 [quant-ph].

\bibitem{Blekos_2024}
K.~Blekos, D.~Brand, A.~Ceschini, C.-H. Chou, R.-H. Li, K.~Pandya, and A.~Summer, ``A review on quantum approximate optimization algorithm and its variants,'' {\em Physics Reports}, vol.~1068, p.~1–66, June 2024.

\bibitem{zhu_adaptive_2020}
L.~Zhu, H.~L. Tang, G.~S. Barron, F.~A. Calderon-Vargas, N.~J. Mayhall, E.~Barnes, and S.~E. Economou, ``An adaptive quantum approximate optimization algorithm for solving combinatorial problems on a quantum computer,'' Dec. 2020.
\newblock arXiv:2005.10258 [quant-ph].

\bibitem{herrman_multi-angle_2022}
R.~Herrman, P.~C. Lotshaw, J.~Ostrowski, T.~S. Humble, and G.~Siopsis, ``Multi-angle quantum approximate optimization algorithm,'' {\em Scientific Reports}, vol.~12, p.~6781, Apr. 2022.

\bibitem{lee_iterative_2023}
X.~Lee, X.~Yan, N.~Xie, Y.~Saito, D.~Cai, and N.~Asai, ``Iterative {Layerwise} {Training} for {Quantum} {Approximate} {Optimization} {Algorithm},'' Sept. 2023.
\newblock arXiv:2309.13552 [quant-ph].

\bibitem{zheng_quantum_2022}
M.~Zheng, B.~Peng, N.~Wiebe, A.~Li, X.~Yang, and K.~Kowalski, ``Quantum algorithms for generator coordinate methods,'' Dec. 2022.
\newblock arXiv:2212.09205.

\bibitem{jamet_quantum_2022}
F.~Jamet, A.~Agarwal, and I.~Rungger, ``Quantum subspace expansion algorithm for {Green}'s functions,'' Dec. 2022.
\newblock arXiv:2205.00094.

\bibitem{beaujeault-taudiere_solving_2023}
Y.~Beaujeault-Taudiere and D.~Lacroix, ``Solving the {Lipkin} model using quantum computers with two qubits only with a hybrid quantum-classical technique based on the {Generator} {Coordinate} {Method},'' Dec. 2023.
\newblock arXiv:2312.04703.

\bibitem{douglas2000introduction}
D.~West, {\em Introduction to Graph Theory (2nd Edition)}.
\newblock Prentice Hall, 08 2000.

\bibitem{gross2018graph}
J.~Gross, J.~Yellen, and M.~Anderson, {\em Graph Theory and Its Applications (3rd ed.)}.
\newblock Chapman and Hall/CRC, 2018.

\bibitem{Henriet_2020}
L.~Henriet, L.~Beguin, A.~Signoles, T.~Lahaye, A.~Browaeys, G.-O. Reymond, and C.~Jurczak, ``Quantum computing with neutral atoms,'' {\em Quantum}, vol.~4, p.~327, Sept. 2020.

\bibitem{Pichler2018}
H.~Pichler, S.-T. Wang, L.~Zhou, S.~Choi, and M.~D. Lukin, ``Quantum optimization for maximum independent set using rydberg atom arrays,'' {\em arXiv preprint arXiv:1808.10816}, 2018.

\bibitem{Kim2021}
M.~Kim, K.~Kim, J.~Hwang, E.-G. Moon, and J.~Ahn, ``Rydberg quantum wires for maximum independent set problems with nonplanar and high-degree graphs,'' {\em arXiv preprint arXiv:2109.03517}, 2021.

\bibitem{Nguyen2023_quantum}
M.-T. Nguyen, J.-G. Liu, J.~Wurtz, M.~D. Lukin, S.-T. Wang, and H.~Pichler, ``Quantum optimization with arbitrary connectivity using rydberg atom arrays,'' {\em PRX Quantum}, vol.~4, p.~010316, Feb 2023.

\bibitem{zhao2024quantumhamiltonianalgorithmsmaximum}
X.~Zhao, P.~Ge, H.~Yu, L.~You, F.~Wilczek, and B.~Wu, ``Quantum hamiltonian algorithms for maximum independent sets,'' 2024.

\bibitem{brady2023iterativequantumalgorithmsmaximum}
L.~T. Brady and S.~Hadfield, ``Iterative quantum algorithms for maximum independent set: A tale of low-depth quantum algorithms,'' 2023.

\bibitem{wybo2024missingpuzzlepiecesperformance}
E.~Wybo and M.~Leib, ``Missing puzzle pieces in the performance landscape of the quantum approximate optimization algorithm,'' 2024.

\bibitem{ni2025progressivequantumalgorithmmaximum}
X.-H. Ni, L.-X. Li, Y.-Q. Song, Z.-P. Jin, S.-J. Qin, and F.~Gao, ``Progressive quantum algorithm for maximum independent set with quantum alternating operator ansatz,'' 2025.

\bibitem{xu2025qaoaparametertransferabilitymaximum}
H.~Xu, X.~Liu, A.~Pothen, and I.~Safro, ``Qaoa parameter transferability for maximum independent set using graph attention networks,'' 2025.

\bibitem{McClean_2018_barren}
J.~R. McClean, S.~Boixo, V.~N. Smelyanskiy, R.~Babbush, and H.~Neven, ``Barren plateaus in quantum neural network training landscapes,'' {\em Nature Communications}, vol.~9, Nov. 2018.

\bibitem{Larocca_2022_diagnosing}
M.~Larocca, P.~Czarnik, K.~Sharma, G.~Muraleedharan, P.~J. Coles, and M.~Cerezo, ``Diagnosing barren plateaus with tools from quantum optimal control,'' {\em Quantum}, vol.~6, p.~824, Sept. 2022.

\bibitem{Larocca_2025_barren}
M.~Larocca, S.~Thanasilp, S.~Wang, K.~Sharma, J.~Biamonte, P.~J. Coles, L.~Cincio, J.~R. McClean, Z.~Holmes, and M.~Cerezo, ``Barren plateaus in variational quantum computing,'' {\em Nature Reviews Physics}, vol.~7, p.~174–189, Mar. 2025.

\bibitem{anand2022exploring}
A.~Anand, S.~Alperin-Lea, A.~Choquette, and A.~Aspuru-Guzik, ``Exploring the role of parameters in variational quantum algorithms,'' 2022.

\bibitem{allcock2024dynamical}
J.~Allcock, M.~Santha, P.~Yuan, and S.~Zhang, ``On the dynamical lie algebras of quantum approximate optimization algorithms,'' 2024.

\bibitem{kazi2024analyzing}
S.~Kazi, M.~Larocca, M.~Farinati, P.~J. Coles, M.~Cerezo, and R.~Zeier, ``Analyzing the quantum approximate optimization algorithm: ans\"atze, symmetries, and lie algebras,'' 2024.

\bibitem{tate_warm-started_2024}
R.~Tate and S.~Eidenbenz, ``Warm-{Started} {QAOA} with {Aligned} {Mixers} {Converges} {Slowly} {Near} the {Poles} of the {Bloch} {Sphere},'' Sept. 2024.
\newblock arXiv:2410.00027.

\bibitem{hill_nuclear_1953}
D.~L. Hill and J.~A. Wheeler, ``Nuclear {Constitution} and the {Interpretation} of {Fission} {Phenomena},'' {\em Physical Review}, vol.~89, pp.~1102--1145, Mar. 1953.

\bibitem{griffin_collective_1957}
J.~J. Griffin and J.~A. Wheeler, ``Collective {Motions} in {Nuclei} by the {Method} of {Generator} {Coordinates},'' {\em Physical Review}, vol.~108, pp.~311--327, Oct. 1957.

\bibitem{Wa_Wong1975-oj}
C.~Wa~Wong, ``Generator-coordinate methods in nuclear physics,'' {\em Phys. Rep.}, vol.~15, pp.~283--357, Jan. 1975.

\bibitem{ring_nuclear_1980}
P.~Ring and P.~Schuck, {\em The {Nuclear} {Many}-{Body} {Problem}}, vol.~103.
\newblock Springer Berlin, Heidelberg, Jan. 1980.

\bibitem{bender_self-consistent_2003}
M.~Bender, P.-H. Heenen, and P.-G. Reinhard, ``Self-consistent mean-field models for nuclear structure,'' {\em Reviews of Modern Physics}, vol.~75, pp.~121--180, Jan. 2003.

\bibitem{verriere_time-dependent_2020}
M.~Verriere and D.~Regnier, ``The time-dependent generator coordinate method in nuclear physics,'' {\em Frontiers in Physics}, vol.~8, p.~233, July 2020.
\newblock arXiv:2004.10147 [nucl-th].

\bibitem{Bertsch_2017}
G.~F. Bertsch, ``The shapes of nuclei,'' {\em International Journal of Modern Physics E}, vol.~26, p.~1740001, Jan. 2017.

\bibitem{Bertsch_2022}
G.~F. Bertsch and K.~Hagino, ``Generator coordinate method for transition-state dynamics in nuclear fission,'' {\em Physical Review C}, vol.~105, Mar. 2022.

\bibitem{severyukhin_beyond_2006}
A.~P. Severyukhin, M.~Bender, and P.-H. Heenen, ``Beyond mean-field study of excited states: {Analysis} within the {Lipkin} model,'' {\em Physical Review C}, vol.~74, p.~024311, Aug. 2006.
\newblock arXiv:nucl-th/0603069.

\bibitem{Nikcic_2011}
T.~Nikšić, D.~Vretenar, and P.~Ring, ``Relativistic nuclear energy density functionals: Mean-field and beyond,'' {\em Progress in Particle and Nuclear Physics}, vol.~66, p.~519–548, July 2011.

\bibitem{bally_symmetry_2012}
B.~Bally, B.~Avez, M.~Bender, and P.-H. Heenen, ``Symmetry restoration for odd-mass nuclei with a {Skyrme} energy density functional,'' {\em International Journal of Modern Physics E}, vol.~21, p.~1250026, May 2012.
\newblock arXiv:1111.0451 [nucl-th].

\bibitem{Egido_2016}
J.~L. Egido, ``State-of-the-art of beyond mean field theories with nuclear density functionals,'' {\em Physica Scripta}, vol.~91, p.~073003, June 2016.

\bibitem{Zhao_2019}
J.~Zhao, J.~Xiang, Z.-P. Li, T.~Nikšić, D.~Vretenar, and S.-G. Zhou, ``Time-dependent generator-coordinate-method study of mass-asymmetric fission of actinides,'' {\em Physical Review C}, vol.~99, May 2019.

\bibitem{jancovici_collective_1964}
B.~Jancovici and D.~H. Schiff, ``The collective vibrations of a many-fermion system,'' {\em Nuclear Physics}, vol.~58, pp.~678--686, Sept. 1964.

\bibitem{Jiao_2019}
C.~Jiao and C.~W. Johnson, ``Union of rotational and vibrational modes in generator-coordinate-type calculations, with application to neutrinoless double-$\beta$ decay,'' {\em Physical Review C}, vol.~100, Sept. 2019.

\bibitem{marevic_quantum_2023}
P.~Marević, D.~Regnier, and D.~Lacroix, ``Quantum fluctuations induce collective multiphonons in finite {Fermi} liquids,'' {\em Physical Review C}, vol.~108, p.~014620, July 2023.
\newblock arXiv:2304.07380 [nucl-th].

\bibitem{umeano2024quantumsubspaceexpansionapproach}
C.~Umeano, F.~Jamet, L.~P. Lindoy, I.~Rungger, and O.~Kyriienko, ``Quantum subspace expansion approach for simulating dynamical response functions of kitaev spin liquids,'' 2024.

\bibitem{zheng_unleashed_2024}
M.~Zheng, B.~Peng, A.~Li, X.~Yang, and K.~Kowalski, ``Unleashed from {Constrained} {Optimization}: {Quantum} {Computing} for {Quantum} {Chemistry} {Employing} {Generator} {Coordinate} {Method},'' Aug. 2024.
\newblock arXiv:2312.07691.

\bibitem{parrish_quantum_2019}
R.~M. Parrish and P.~L. McMahon, ``Quantum {Filter} {Diagonalization}: {Quantum} {Eigendecomposition} without {Full} {Quantum} {Phase} {Estimation},'' Sept. 2019.
\newblock arXiv:1909.08925 [quant-ph].

\bibitem{stair2019multireferencequantumkrylovalgorithm}
N.~H. Stair, R.~Huang, and F.~A. Evangelista, ``A multireference quantum krylov algorithm for strongly correlated electrons,'' 2019.

\bibitem{lee2021variationalquantumsimulationchemical}
C.-K. Lee, C.-Y. Hsieh, S.~Zhang, and L.~Shi, ``Variational quantum simulation of chemical dynamics with quantum computers,'' 2021.

\bibitem{marev2018}
P.~Marević, {\em Towards a unified description of quantum liquid and cluster states in atomic nuclei within the relativistic energy density functional framework}.
\newblock PhD thesis, Université Paris-Saclay, 2018.
\newblock Thèse de doctorat dirigée par Khan, Elias Structure et réactions nucléaires Université Paris-Saclay (ComUE) 2018.

\bibitem{hamermesh_group_1962}
M.~M. Hamermesh, {\em Group theory and its application to physical problems}.
\newblock Reading, MA. : Addison-Wesley Pub. Co., 1962.

\bibitem{lipkin_lie_1966}
H.~J. Lipkin, {\em Lie groups for pedestrians}.
\newblock Amsterdam : North-Holland Pub. Co. [sole distributors for U.S.A. and Canada, Inter-science Publishers, New York], 1966.

\bibitem{matsumoto_extension_2023}
M.~Matsumoto, Y.~Tanimura, and K.~Hagino, ``An extension of the generator coordinate method with basis optimization,'' {\em Physical Review C}, vol.~108, p.~L051302, Nov. 2023.
\newblock arXiv:2308.13233 [nucl-th].

\bibitem{Ritz1909}
W.~Ritz, ``Über eine neue methode zur lösung gewisser variationsprobleme der mathematischen physik,'' {\em Journal für die reine und angewandte Mathematik}, 1909.

\bibitem{bonche_analysis_1990}
P.~Bonche, J.~Dobaczewski, H.~Flocard, P.~H. Heenen, and J.~Meyer, ``Analysis of the generator coordinate method in a study of shape isomerism in {194Hg},'' {\em Nuclear Physics A}, vol.~510, pp.~466--502, Apr. 1990.

\bibitem{epperly_theory_2023}
E.~N. Epperly, L.~Lin, and Y.~Nakatsukasa, ``A theory of quantum subspace diagonalization,'' June 2023.
\newblock arXiv:2110.07492.

\bibitem{higgott_variational_2019}
O.~Higgott, D.~Wang, and S.~Brierley, ``Variational {Quantum} {Computation} of {Excited} {States},'' {\em Quantum}, vol.~3, p.~156, July 2019.

\bibitem{childs_hamiltonian_2012}
A.~M. Childs and N.~Wiebe, ``Hamiltonian {Simulation} {Using} {Linear} {Combinations} of {Unitary} {Operations},'' {\em Quantum Information and Computation}, vol.~12, Feb 2012.
\newblock arXiv:1202.5822 [quant-ph].

\bibitem{Byrd1995_Broyden}
R.~H. Byrd, P.~Lu, J.~Nocedal, and C.~Zhu, ``A limited memory algorithm for bound constrained optimization,'' {\em SIAM Journal on Scientific Computing}, vol.~16, no.~5, pp.~1190--1208, 1995.

\bibitem{Zhu_1997_Broyden}
C.~Zhu, R.~H. Byrd, P.~Lu, and J.~Nocedal, ``Algorithm 778: L-bfgs-b: Fortran subroutines for large-scale bound-constrained optimization,'' {\em ACM Trans. Math. Softw.}, vol.~23, p.~550–560, Dec. 1997.

\bibitem{cazals_identifying_2025}
P.~Cazals, A.~François, L.~Henriet, L.~Leclerc, M.~Marin, Y.~Naghmouchi, W.~d.~S. Coelho, F.~Sikora, V.~Vitale, R.~Watrigant, M.~W. Garzillo, and C.~Dalyac, ``Identifying hard native instances for the maximum independent set problem on neutral atoms quantum processors,'' Feb. 2025.
\newblock arXiv:2502.04291 [quant-ph].

\bibitem{wiki:Finite_difference_coefficient}
Wikipedia, ``{Finite difference coefficient} --- {W}ikipedia{,} the free encyclopedia.'' \url{http://en.wikipedia.org/w/index.php?title=Finite\%20difference\%20coefficient&oldid=1275142675}, 2025.
\newblock [Online; accessed 20-March-2025].

\bibitem{ross2016optimalancillafreecliffordtapproximation}
N.~J. Ross and P.~Selinger, ``Optimal ancilla-free clifford+t approximation of z-rotations,'' 2016.

\bibitem{Barenco_1995}
A.~Barenco, C.~H. Bennett, R.~Cleve, D.~P. DiVincenzo, N.~Margolus, P.~Shor, T.~Sleator, J.~A. Smolin, and H.~Weinfurter, ``Elementary gates for quantum computation,'' {\em Physical Review A}, vol.~52, p.~3457–3467, Nov. 1995.

\bibitem{Nielsen_Chuang10}
M.~A. Nielsen and I.~L. Chuang, {\em Quantum Computation and Quantum Information: 10th Anniversary Edition}.
\newblock Cambridge University Press, 2010.

\bibitem{Gidney_2015_Constructing}
C.~Gidney, ``{C}onstructing {L}arge {C}ontrolled {N}ots --- algassert.com.'' \url{https://algassert.com/circuits/2015/06/05/Constructing-Large-Controlled-Nots.html}.
\newblock [Accessed 16-05-2025].

\bibitem{Claudon_2024}
B.~Claudon, J.~Zylberman, C.~Feniou, F.~Debbasch, A.~Peruzzo, and J.-P. Piquemal, ``Polylogarithmic-depth controlled-not gates without ancilla qubits,'' {\em Nature Communications}, vol.~15, July 2024.

\end{thebibliography}
